# Kramers equation algorithm for simulations of QCD with two flavors of Wilson fermions and gauge group SU(2)

Karl Jansen and Chuan Liu

Deutsches Elektronen-Synchrotron DESY,

Notkestr. 85, D-22603 Hamburg, Germany

June 13, 1995

## Abstract

We compare the Hybrid Monte Carlo (HMC) and the Kramers equation algorithms for simulations of QCD with two flavors of dynamical Wilson fermions and gauge group $SU(2)$. The results for the performance of both algorithms are obtained on $6^3 12$, $12^4$ and $16^4$ lattices at a pion to $\rho$ meson mass ratio of $m_\pi/m_\rho \approx 0.9$. We find that the Kramers equation algorithm gives an equally good performance as the HMC algorithm. We demonstrate that the classical equations of motion used in these algorithms lack reversibility in practical simulations and behave like those of a chaotic dynamical system with a Liapunov exponent $\nu \approx 0.75$.

## 1 Introduction

Numerical simulations have been proven to be an important tool in obtaining information about non-perturbative properties of a physical system. Finding efficient algorithms is therefore one of the major research subjects. In particular, improved algorithms for models containing fermions, notably QCD, are needed. Up to now, the Hybrid Monte Carlo (HMC) algorithm [1] has been the most often used update scheme for fermionic systems. It is an exact Monte Carlo method, easily implementable on vector and parallel machines and it has been proven to work well in practice. Still, simulations of fermionic systems turn out to be very difficult and time consuming. Improvements on this situation are therefore clearly welcome and alternatives to the HMC algorithm should be tried out.

Recently, a new idea has been put forward by M. Lüscher [2] that establishes another exact method for QCD simulations. It consists of a "bosonization" of the fermionic system. One arrives at a completely local bosonic action. The price to pay is the introduction of a number of scalar field copies, where the number of copies typically is $O(100)$ on a $16^4$ lattice. Unfortunately, unexpected long autocorrelation times have been encountered in practical simulations within a QCD like theory [3, 4], where the gauge group was chosen to be SU(2) instead of SU(3). So far, it remains to be seen whether the problem of these large autocorrelation times can be overcome.

In this paper, we want to continue the search for fermion algorithms by studying the Kramers equation algorithm. This algorithm was proposed by Horowitz [5] already some time ago and is a generalization of the HMC algorithm. The main modification is that, in the refreshment of the momenta, a term proportional to the momenta themselves is added. It can be made exact by introducing a global accept/reject step. At least for a free field theory, it can be shown to have a dynamical critical exponent of $z = 1$. However, in contrast to the HMC algorithm, this result can be obtained without going to the large trajectory length limit. The hope is therefore that, as far as the critical slowing down is concerned, it works equally well as the HMC algorithm. However, due to the shorter trajectory lengths, it might consume less computer time. The Kramers equation algorithm was introduced and discussed in [5]. A particular implementation and further discussion can be found in [6], where a test of the algorithm was performed on the 2-dimensional Gross-Neveu model. The results in [6] indicate that the Kramers equation performs as well as the HMC algorithm. A test of the Kramers equation algorithm for QCD has not been performed so far.

Here we want to fill this gap and study the Kramers equation algorithm in QCD with $SU(2)$ gauge group. The reason why we have chosen SU(2) instead of SU(3) is that we want to continue the algorithm tests as initiated in [3, 4] with Lüscher's fermion algorithm. Our results for Wilson QCD show that, in the actual computer time, the Kramers equation algorithm performs at least as well as the conventional Hybrid Monte Carlo algorithm.

Another aspect which is in favor of the Kramers equation algorithm, is the lack of reversibility of the discretized classical equations of motion in the HMC algorithm. Since for lattices used in todays simulations, the number of steps within a trajectory can reach $O(100)$ [7], one may fear that due to accumulations of rounding errors, the HMC algorithm lacks its reversibility property, necessary for the detailed balance condition. Indeed, in numerical experiments violations of reversibility have been observed [8]. In the Kramers equation algorithm, this effect is drastically reduced since only one step in the leapfrog integration is used.

This paper is organized as follows. After defining our model in the next paragraph, we recapitulate shortly the HMC algorithm in section 2. In section 3, we introduce the Kramers equation algorithm. Two improvements we have used, namely preconditioning and a better leapfrog integration scheme as proposed in [9] are explained in section 4. The performance tests of the Kramers equation against the HMC algorithm are presented in section 5. Section 6 is devoted to the investigation of reversibility in the HMC algorithm and we will conclude in section 7.

The theory that we would like to study is the standard lattice Wilson QCD with gauge group $SU(2)$. We will work on a 4-dimensional euclidean space-time lattice with volume $\Omega = L_s^3 L_t$. We introduce gauge fields $U_\mu(x) \in SU(2)$ where $\mu = 0, 1, 2, 3$ designates the 4 forward directions in space-time and quark fields $\psi_{Aa\alpha}(x)$ where $A$,$a$ and $\alpha$ are flavor, color and Dirac indices, respectively. The full partition function for our model is given by,

$$\mathcal{Z} = \int \mathcal{D}U \mathcal{D}\bar{\psi} \mathcal{D}\psi \exp\left(-S_g - S_w\right), \tag{1}$$

where the gauge action $S_g$ and the Wilson fermion action $S_w$ are given by:

$$\begin{align}
S_g &= -\frac{\beta}{2} \sum_P Tr(U_P), \\
S_w &= \sum_x \bar{\psi}(x)(D + m)\psi(x).
\end{align} \tag{2}$$

The Wilson difference operator $D$ which appears in the above expression is given by (setting the Wilson parameter to one):

$$\begin{align}
D &= \frac{1}{2} \sum_\mu \gamma_\mu (\nabla_\mu + \nabla_\mu^*) - \nabla_\mu \nabla_\mu^*, \\
\nabla_\mu \psi(x) &= U_\mu(x)\psi(x + \mu) - \psi(x), \\
\nabla_\mu^* \psi(x) &= \psi(x) - U_\mu^\dagger(x - \mu)\psi(x - \mu),
\end{align} \tag{3}$$

and $U_P$ is the usual plaquette term on the lattice. In the following, we will consider two degenerate flavors of Wilson fermions. As usual, for the simulations the fermion determinant is written in terms of Gaussian scalar fields $\phi$ such that the path integral reads

$$\begin{align}
\mathcal{Z} &= \int \mathcal{D}U \mathcal{D}\phi^\dagger \mathcal{D}\phi \, e^{-S_{eff}}, \\
S_{eff} &= S_g + \phi^\dagger (M^\dagger M)^{-1} \phi,
\end{align} \tag{4}$$

with $M = 2\kappa(D + m)$ the fermion matrix. The Wilson hopping parameter $\kappa$ is related to the bare quark mass via $\kappa = (8 + 2m)^{-1}$.

## 2  Hybrid Monte Carlo Algorithm

Let $q$ represent some stochastic variable, i.e. the gauge link fields and the scalar fields of eq. (4) in our case. Let us furthermore introduce a fictitious Monte Carlo time $t$ and a set of momenta $p$ which are conjugate to $q$. In the HMC algorithm, the system defined by the euclidean action $S(q)$ evolves then according to (stochastic) Hamilton's equations of motion which read, considering time to be continuous for the moment,

$$\begin{aligned} \dot{p} &= -\frac{\delta H}{\delta q} , \\ \dot{q} &= \frac{\delta H}{\delta p} . \end{aligned} \qquad (5)$$

Here the Hamiltonian $H = \frac{1}{2}p^2 + S(q)$ and the initial momenta are obtained from random numbers with Gaussian measure of unit variance. A leapfrog integrator [1] with $N_{md}$ molecular dynamics steps and a discrete step size $\epsilon$ is used to integrate eqs. (5) numerically. The product $\epsilon N_{md}$ is called the trajectory length and a trajectory consists of $N_{md}$ molecular dynamics steps. In the simulations $\epsilon N_{md} = O(1)$.

In the HMC algorithm, due to the finite step size errors of the leapfrog integration, a global accept/reject step at the end of a trajectory is needed to ensure exactness of the algorithm. In a free theory, one can show that the acceptance rate of the Hybrid Monte Carlo algorithm is given by [10, 11]:

$$P_{acc} \sim \text{erfc}(cN_{md}\epsilon^3\sqrt{\Omega}) , \qquad (6)$$

where erfc is the error function and $c$ a constant. If the total trajectory length $\epsilon N_{md}$ is fixed, an increasing $\epsilon$ will drive the acceptance rate to zero exponentially. Therefore, one should take as large a step size as possible, while keeping the acceptance rate at a reasonably high level. To maintain a constant acceptance rate, one also has to scale the step size according to $\Omega^{-1/4}$. In our simulations, we have tried to maintain the acceptance rate at 80 to 90 percent level.

## 3  Kramers equation algorithm

The Kramers equation algorithm for simulating quantum field theories was introduced and described in [5]. It amounts to add a second order time derivative term to the usual Langevin equation and originates from the theory of Brownian motion (see [12] for a discussion of the Kramers equation in this context). The equations describing the time evolution of the system are very similar to eqs. (5) and read

$$\begin{aligned} \dot{p} &= -\frac{\delta H}{\delta q} - \gamma p + \eta(t) , \\ \dot{q} &= \frac{\delta H}{\delta p} , \end{aligned} \qquad (7)$$

where the stochastic variables $\eta(t)$ are the so-called "white noise". Of course, eqs. (7) are to be understood only on a formal level. We see that the main difference between eqs. (5) and eqs. (7) is the white noise term and an addition of a friction term which introduces a new tunable parameter $\gamma$.

In this paper, we adopt a particular variant of the Kramers equation algorithm [5, 6] which in its discretized version consists of the following steps.

1.) Generate momenta $p_1$ according to a Gaussian distribution with zero mean and unit variance

$$P(p_1) \propto e^{-\frac{p_1^2}{2}} . \qquad (8)$$

Given the initial configuration $(q_1, p_1)$,

2.) update the momenta, using a discretized time step $\epsilon$

$$p = e^{-\gamma\epsilon} p_1 + \sqrt{1 - e^{-2\gamma\epsilon}} \eta , \qquad (9)$$

where $\eta$ is again selected from a Gaussian distribution with zero mean and unit variance.

3.) Leapfrog integration

$$\begin{aligned} \tilde{p} &= p + \frac{\epsilon}{2} F(q_1) , \\ q_2 &= q_1 + \epsilon \tilde{p} , \\ p_2 &= \tilde{p} + \frac{\epsilon}{2} F(q_2) , \end{aligned} \qquad (10)$$

where $F = -\delta S/\delta q$ is the force.

4.) Perform a Metropolis test, by accepting the candidate configuration with probability

$$P(q_1, p \to q_2, p_2) = \min\{1, e^{H(q_1,p) - H(q_2,p_2)}\} . \qquad (11)$$

On rejection, set

$$q_2 = q_1 , \quad p_2 = -p , \qquad (12)$$

where the negation of the momenta is necessary to guarantee exactness of the algorithm. In the appendix, we show that the above scheme fulfills the stability criterion, which –in combination with ergodicity– ensures the convergence to the ground state probability distribution [13]. In the simulations, steps 2.)-4.) are repeated $k$-times before the momenta are refreshed again in step 1.). In our simulations, a value of $k = 4$ was chosen. Obviously, in the limit $\gamma = \infty$ and $k = 1$, the above scheme reduces to a one step Hybrid Monte Carlo algorithm.

A continuum free field analysis of eqs. (7) [5, 6] reveals that the exponential autocorrelation time behaves as $1/\omega_{min}$, where $\omega_{min}$ is the slowest mode of the system, i.e. the inverse correlation length $\xi$. This behavior, corresponding to a dynamical critical exponent of $z = 1$ as in the HMC algorithm, is assumed at $\gamma = 2\omega_{min}$.

The remarkable property of the Kramers equation algorithm is that this result can be obtained already for short trajectory lengths, whereas for the HMC algorithm, a value of $z = 1$ can only be reached in the large trajectory length limit [11, 14]. This property of the Kramers equation algorithm has important consequences. First, one obviously saves computer time per trajectory. Second, since we have only one step in the trajectory, effects of the (non-)reversibility in the HMC algorithm (see Section 6) are drastically reduced. Third, according to the free field analysis of the tuning of the step size (6), the step size is expected to scale like $\epsilon \propto \Omega^{-1/6}$. This softer volume dependence of the step size is indeed seen in our practical simulations.

The drawback, of course, is that the trajectory lengths become shorter, resulting in larger autocorrelation times. Whether the advantages mentioned above will merit the increase of the autocorrelation times, will be investigated in section 5, where we give results for the performance of the HMC and the Kramers equation algorithms in simulations of QCD with gauge group $SU(2)$. In the next section, we will first explain two improvements that we have implemented for both the HMC and the Kramers equation algorithms.

## 4 Improvements

**Even-odd Preconditioning**

Preconditioning [15] is by now standard for simulations in QCD. We write the fermion matrix as

$$M \equiv \begin{pmatrix} 1 & -\kappa D_{eo} \\ -\kappa D_{oe} & 1 \end{pmatrix}. \quad (13)$$

The nondiagonal part of the fermion matrix $D_{eo}$ only connects the odd lattice points with even lattice points and similarly the matrix $D_{oe}$ only connects the even lattice points to the odd lattice points. The preconditioned matrix $\tilde{M}$ is now

$$\tilde{M} = \begin{pmatrix} 1 & 0 \\ 0 & 1 - \kappa^2 D_{oe} D_{eo} \end{pmatrix}, \quad (14)$$

and the path integral (4) can be written equally in terms of the preconditioned matrix $\tilde{M}$.

The preconditioned matrix has two advantages over the original fermion matrix. The first is a reduction of the memory requirement. Since $\tilde{M}$ only connects odd with odd (or even with even) sites, we save a factor of two in the memory requirement. The second advantage is that the matrix $\tilde{M}^\dagger \tilde{M}$, which is used in the simulations, is better conditioned than the original matrix $M^\dagger M$.

Let us denote $\lambda_{max}$ and $\lambda_0$ to be the largest and the lowest eigenvalue of the matrix $M^\dagger M$ and $\tilde{\lambda}_{max}, \tilde{\lambda}_0$ will correspond to the largest and the lowest eigenvalue of the matrix $\tilde{M}^\dagger \tilde{M}$. One can show that

$$\sqrt{\lambda_{max}} = ||M|| \leq 1 + \kappa ||D|| \leq 1 + 8\kappa, \quad (15)$$

and similarly $\sqrt{\tilde{\lambda}_{max}} = ||\tilde{M}|| \leq 1 + 64\kappa^2$, where the notation $||.||$ stands for the $l_2$-norm of the matrix [16]. The ratio $(1 + 64\kappa^2)/(1 + 8\kappa)$ is close to one for all practical values of $\kappa$ in numerical simulations. Therefore, when comparing the condition number $\lambda_{max}/\lambda_0$ of the matrix $M^\dagger M$ with $\tilde{\lambda}_{max}/\tilde{\lambda}_0$ of the preconditioned matrix $\tilde{M}^\dagger \tilde{M}$, the main difference comes from the ratio of the lowest eigenvalue in the two cases, assuming that the largest eigenvalue is close to the bound given above. It is expected that the lowest eigenvalue $\tilde{\lambda}_0$ of the preconditioned matrix $\tilde{M}^\dagger \tilde{M}$ is about a factor of 4 larger than the lowest eigenvalue of the original matrix $M^\dagger M$. This is motivated by noticing that for any eigenvalue $\lambda$ of the matrix $M$, $(2\lambda - \lambda^2)$ is an eigenvalue of the matrix $\tilde{M}$. Therefore, when $|\lambda|$ is small, the absolute value of the corresponding eigenvalue of $\tilde{M}$ is about a factor of two larger than that of $M$. However, since the eigenvalues of $M$ are not directly related to the eigenvalues of $M^\dagger M$, we have to test this expectation numerically. Indeed, the effect is confirmed by our numerical simulation results. Since the number of Conjugate Gradient iterations is proportional to the square root of the condition number, we expect to save about a factor of two in computer time.

Table 1: Comparison of the average lowest eigenvalue for the original matrix $M^\dagger M$ ($<\lambda_0>$) and the preconditioned matrix $\tilde{M}^\dagger \tilde{M}$ ($<\tilde{\lambda}_0>$). $N_{CG}$ is the average number of Conjugate Gradient iterations to reach a given residue and $<U_P>$ is the average plaquette value. The subscript *old* stands for the original and *pre* for the preconditioned matrix.

| $\kappa$ | $<\lambda_0>$ | $<U_{P,old}>$ | $N_{CG,old}$ | $<\tilde{\lambda}_0>$ | $<U_{P,pre}>$ | $N_{CG,pre}$ |
|---|---|---|---|---|---|---|
| 0.150 | 0.3650(103) | 0.449(2) | 57 | 1.267(22) | 0.4469(3) | 35 |
| 0.160 | 0.1960(20) | 0.461(2) | 76 | 0.7520(10) | 0.4607(4) | 46 |
| 0.170 | 0.1210(50) | 0.490(2) | 95 | 0.3893(87) | 0.4954(4) | 54 |

We have tested the even-odd preconditioned version of the HMC algorithm on a small lattice ($4^4$) for various $\kappa$ values at $\beta = 1.75$ and compared with the version using the original matrix $M^\dagger M$. The results are summarized in table 1. The lowest eigenvalue $\lambda_0$ ($\tilde{\lambda}_0$) of the matrix $M^\dagger M$ ($\tilde{M}^\dagger \tilde{M}$) was measured by minimizing the Ritz functional, $\mu(\psi) = ||M\psi||^2/||\psi||^2$, using a Conjugate Gradient technique [17, 18]. One sees that indeed the average value of the lowest eigenvalue $<\tilde{\lambda}_0>$ of $\tilde{M}^\dagger \tilde{M}$ is always about a factor of 3.5 larger than $<\lambda_0>$ of the original matrix $M^\dagger M$. Correspondingly, the number of Conjugate gradient iterations $N_{CG}$ for each inversion of the matrix is decreased. The average plaquette values $<U_P>$ in both cases are also listed for comparison. A similar improvement was also found in the larger volume simulations. For the lattice size of $6^3 12$ and $\beta = 2.12$, $\kappa = 0.15$, we found the average value $<\tilde{\lambda}_0>$ to be larger by about a factor of 3.6 as compared to $<\lambda_0>$.

**Sexton-Weingarten Integration**

Another improvement can be made to the leapfrog integration scheme used in the conventional Hybrid Monte Carlo algorithm. This has been suggested by Sexton and Weingarten in [9]. The basic idea is the following. The force $F$ needed for the update of the momenta consists of two parts. One originates from the gauge staples and the other from the fermionic force. While the pure gauge part needs only moderate computer time, the fermionic force part includes a matrix inversion and consumes considerably more CPU time.

One may, however, introduce different time scales for the leapfrog integration corresponding to the two force terms. For the pure gauge part, the leapfrog can be done with finer time steps. Although we then have to do more arithmetic operations for this part, we finally gain in the increase of the acceptance rate due to partial cancellations of the finite step size errors coming from the force.

Consider the Hamiltonian

$$H = \frac{1}{2}p^2 + S_1(q) + S_2(q) , \qquad (16)$$

for which the number of arithmetic operations required to evaluate the force due to $S_1$ is far less than that due to $S_2$. Then, the partial time evolution operators are given by

$$\begin{aligned} T_1(\epsilon) &= \exp(\frac{1}{2}\epsilon L(S_1)) \exp(\epsilon L(\frac{1}{2}p^2)) \exp(\frac{1}{2}\epsilon L(S_1)) , \\ T_2(\epsilon) &= \exp(\epsilon L(S_2)) . \end{aligned} \qquad (17)$$

In eq. (17) $L$ is a linear operator, representing a symplectic integrator [9, 6]. Sexton and Weingarten suggest that one can define a full time evolution operator $T(\epsilon)$ by:

$$T(\epsilon) = T_2(\frac{\epsilon}{2}) \left[T_1(\frac{\epsilon}{n})\right]^n T_2(\frac{\epsilon}{2}) . \qquad (18)$$

As has been shown in [9], errors induced by finite time step sizes will be reduced by the above method due to finer step sizes, leading to a better integration scheme. A more complicated scheme can be constructed by adding one more $T_2$ insertion in each step. The scheme we used is given by the following formal expression:

$$\begin{aligned} T(\epsilon) &= T_2(\frac{\epsilon}{6}) \left[(\hat{T}_1(\frac{\epsilon}{2})T_2(\frac{2\epsilon}{3})\hat{T}_1(\frac{\epsilon}{2})T_2(\frac{\epsilon}{3})\right]^{N_{md}-1} \hat{T}_1(\frac{\epsilon}{2})T_2(\frac{2\epsilon}{3})\hat{T}_1(\frac{\epsilon}{2})T_2(\frac{\epsilon}{6}) , \\ \hat{T}_1(\frac{\epsilon}{2}) &= T_g(\frac{\epsilon}{12n}) \left[T_k(\frac{\epsilon}{4n})T_g(\frac{\epsilon}{3n})T_k(\frac{\epsilon}{4n})T_g(\frac{\epsilon}{6n})\right]^{n-1} T_k(\frac{\epsilon}{4n})T_g(\frac{\epsilon}{3n})T_k(\frac{\epsilon}{4n})T_g(\frac{\epsilon}{12n}) . \end{aligned} \qquad (19)$$

In the above formula, a factor of $T_2(\epsilon)$ stands for an update of the momenta by a step $\epsilon$, taking into account the force coming from the fermionic part only. Similarly, $T_g(\epsilon)$ stands for an update of the momenta by a step $\epsilon$ due to the gauge part alone and $T_k(\epsilon)$ stands for an update of the gauge fields.

In the practical simulation, the integer $n$ is chosen to be around 4. Increasing this number will not result in a further improvement of the acceptance rate [9]. The total trajectory length in such a sequence of updates is $t = \epsilon N_{md}$ and the number of matrix inversions is equal to $(2N_{md} + 1)$. More complicated schemes than the above will require more $T_2$ insertions in each step and will not lead to further improvements.

The tests of the above scheme as given in eq. (19) on the small lattice ($4^4$) runs turned out to be quite promising. In table 2, we have listed the results of the tests for the conventional integration scheme and the Sexton-Weingarten scheme for the HMC algorithm. These simulations were done on a $4^4$ lattice with $\beta = 1.75$ and $\kappa = 0.15$. We fixed the trajectory length to be one in all these runs, which is a reasonable value for the practical runs as well. Then, we systematically changed the number of steps of the trajectory hence the step size. The number of conjugate gradient iterations needed for each trajectory ($N_{iter}$) and the acceptance rate ($P_{acc}$) of the runs are listed in table 2. The ratio $N_{iter}/P_{acc}$ should be minimized for optimal performance. From table 2 we see that, for the conventional leapfrog integration scheme, the best performance is reached at about 80 percent acceptance rate. The corresponding value of $N_{iter}/P_{acc}$ is about 420. The Sexton-Weingarten integration scheme, however, can achieve a $N_{iter}/P_{acc}$ ratio of about 270, which is about 50 percent better than the conventional scheme. Using this new integration scheme, we can increase the step size quite substantially. The effect of these improvements on larger lattices is expected to be more pronounced [9].

Table 2: Comparison of the Sexton-Weingarten integration scheme with the conventional leapfrog scheme. The lattice size is $4^4$ with $\beta = 1.75$ and $\kappa = 0.15$. The total trajectory length has been fixed to one.

| Conventional Leap-frog Integration Scheme | | | | |
|---|---|---|---|---|
| $\epsilon$ | $N_{md}$ | $N_{iter}$ | $P_{acc}$ | $N_{iter}/P_{acc}$ |
| 0.05 | 20 | 691(2) | 0.97(1) | 712(07) |
| 0.0625 | 16 | 576(5) | 0.95(1) | 608(10) |
| 0.0833 | 12 | 454(5) | 0.91(1) | 499(10) |
| 0.10 | 10 | 389(5) | 0.88(2) | 442(15) |
| 0.111 | 9 | 362(2) | 0.83(2) | 437(10) |
| 0.125 | 8 | 329(4) | 0.78(4) | 421(20) |
| 0.167 | 6 | 265(2) | 0.62(2) | 427(20) |
| 0.20 | 5 | 235(4) | 0.52(3) | 451(25) |
| Sexton-Weingarten Integration Scheme | | | | |
| 0.10 | 10 | 735(10) | 1.00(1) | 735(10) |
| 0.125 | 8 | 606(10) | 1.00(1) | 606(10) |
| 0.167 | 6 | 472(08) | 1.00(1) | 472(08) |
| 0.20 | 5 | 405(05) | 0.99(2) | 410(08) |
| 0.25 | 4 | 331(03) | 0.97(3) | 341(10) |
| 0.333 | 3 | 262(03) | 0.95(3) | 275(10) |
| 0.50 | 2 | 195(02) | 0.66(2) | 295(12) |

Table 3: Average plaquette value $<U_p>$, average lowest eigenvalue $<\tilde{\lambda}_0>$ of $\tilde{M}^\dagger\tilde{M}$, the pion and $\rho$ meson masses at $\kappa = 0.15$, $\beta = 2.12$ for $6^3 12$ and $12^4$ lattices. The statistics is given in terms of *measured* trajectories. For the Kramers equation algorithm only every sixth (fourth) trajectory was measured on the $6^3 12$ ($12^4$) lattice. For the HMC algorithm, every trajectory was measured.

| Algorithm | $L_s^3 L_t$ | Trajectories | $<U_p>$ | $<\tilde{\lambda}_0>$ | $m_\pi$ | $m_\rho$ |
|---|---|---|---|---|---|---|
| Kramers | $6^3 12$ | 1060 | 0.5803(2) | 0.0251(2) | 1.225(8) | 1.317(9) |
| HMC | $6^3 12$ | 2080 | 0.5800(2) | 0.0253(2) | 1.213(7) | 1.299(9) |
| Kramers | $12^4$ | 1695 | 0.5779(3) | 0.01472(13) | 1.030(7) | 1.098(10) |
| HMC | $12^4$ | 1060 | 0.5777(3) | 0.01492(10) | 1.047(10) | 1.113(10) |

## 5  Performance tests

In the implementation of the HMC and Kramers equation algorithms, we adopted the $\Phi$ algorithm [19]. The simulations were performed on the Quadrics (APE) machines at DESY, Q1 with 8 nodes and QH2 with 256 nodes. Since these machines have only 32 bit precision, we used a Kahan summation technique [20, 21] to achieve accurate results for scalar products and other global summations. We also tried the biconjugate gradient method. However, similar to the conclusions in [6], we did not find an improvement on the APE computer.

As mentioned in the introduction, we want to continue the search for an alternative to and hopefully better algorithm than the HMC algorithm. We therefore performed our tests at the same parameter values as chosen in [3, 4], i.e. $\beta = 2.12$, $\kappa = 0.15$ and lattice sizes of $6^3 12$, $12^4$ and $16^4$. For the lattice size $16^4$, we only ran the Kramers equation algorithm. We have measured the average plaquette value $<U_p>$, the average lowest eigenvalue $<\tilde{\lambda}_0>$ of the preconditioned matrix $\tilde{M}^\dagger\tilde{M}$, the pion correlation functions and the $\rho$ meson correlation functions in our runs. When we give values of the pion mass $m_\pi$ and the $\rho$ meson mass $m_\rho$, we use as a definition

$$\cosh m = \frac{C(L_t/2 + 1)}{C(L_t/2)}, \qquad (20)$$

where $C$ is the correlation function for the pion or $\rho$ meson. At the chosen values of $\beta$ and $\kappa$, we obtain a ratio of $m_\pi/m_\rho \approx 0.9$.

In order to get an estimate for the performance of the Kramers equation and the HMC algorithms, we have studied the integrated autocorrelation times of the plaquette $U_p$, the lowest eigenvalue $\tilde{\lambda}_0$ of the preconditioned matrix $\tilde{M}^\dagger\tilde{M}$, the pion $C_\pi$ and the $\rho$ meson $C_\rho$ correlation functions at distance $L_t/2 + 1$. We list the values of $<U_p>$, $<\tilde{\lambda}_0>$, $m_\pi$ as well as $m_\rho$ in table 3 for both algorithms, demonstrating that they give consistent results.

In an ideal situation, the integrated autocorrelation time $\tau_{int}$ of some observable $\mathcal{O}$ is ob-

tained via
$$\tau_{int} = \frac{1}{2} \sum_{t=-\infty}^{+\infty} \Gamma_{\mathcal{O}}(t)/\Gamma_{\mathcal{O}}(0), \qquad (21)$$

where $\Gamma_{\mathcal{O}}(t)$ is the connected autocorrelation function of the observable $\mathcal{O}$. In reality, the summation can, of course, only be taken over a finite data set. Summing over the whole data set, assuming that the statistics is much larger than the autocorrelation time, may also be misleading, since the noise in the tail of the autocorrelation functions gives an unwanted bias. Therefore, we have used the "window" method suggested by Sokal [13] to extract the integrated autocorrelation times. This method basically suggests that one should only sum the autocorrelation function up to a certain distance $T_{cut}$, thereby introducing the integrated autocorrelation function
$$\tau_{int}(T_{cut}) = \frac{1}{2} + \sum_{t=1}^{T_{cut}} \Gamma_{\mathcal{O}}(t)/\Gamma_{\mathcal{O}}(0). \qquad (22)$$

In practice, the integrated autocorrelation time of some observable $\mathcal{O}$, $\bar{\tau}_{int}(\mathcal{O})$, is determined by searching for a plateau behavior of the corresponding integrated autocorrelation function in the range of $T_{cut}/\bar{\tau}_{int}(\mathcal{O}) \sim 4 - 10$. The cut parameter $T_{cut}$ should also satisfy $T_{total}/T_{cut} \gg 1$, where $T_{total}$ is the total number of measurements. In Fig. 1, we plot the integrated autocorrelation function for $\tilde{\lambda}_0$ to give an example. One indeed observes a plateau behavior where the integrated autocorrelation function does not depend on the value of $T_{cut}$.

In table 4, we give the results for the integrated ($\bar{\tau}_{int}$) and exponential ($\tau_{exp}$) autocorrelation times for the plaquette value $U_p$ and the pion correlation function $C_\pi$ at distance $L_t/2 + 1$ *measured per trajectory*. For the runs on the $6^3 12$ lattice, we ran eight copies of the program on single nodes of the APE Q1 machine. This allowed us to determine the integrated autocorrelation times for each copy independently, from which we obtained the error quoted in table 4. For the $12^4$ and $16^4$ lattices, the error was obtained by splitting the total run into smaller parts and analyze these parts independently.

We checked the values of the integrated autocorrelation times as given in table 4 by a blocking analysis of the error of the observables. Our statistics is sufficiently large that we can reach an error plateau. The integrated autocorrelation times determined in this way are in complete agreement with the ones obtained from the integrated autocorrelation functions. In addition, we determined the exponential autocorrelation times $\tau_{exp}$ by fitting the time behavior of the autocorrelation functions to an exponential. Again, $\tau_{exp} \approx \bar{\tau}_{int}$ (see table 4) in all cases.

In the last column of table 4, we give the product of the integrated autocorrelation time $\bar{\tau}_{int}(U_p)$ for the plaquette value and the average number of Conjugate Gradient iterations per trajectory. This quantity gives a direct estimate for the computer time consumption of both algorithms and hence should be used for comparison.

Two comments are in order concerning table 4. First, comparing the results for the autocorrelation times on the $6^3 12$ and the $12^4$ lattices, one notices $\bar{\tau}_{int}(U_p)$ to be smaller on the larger lattice. This demonstrates that, by a more careful tuning of the parameters, substantial improvements can be achieved. As we started our investigation with the $6^3 12$ lattice, we did not choose the optimal values for the parameters.

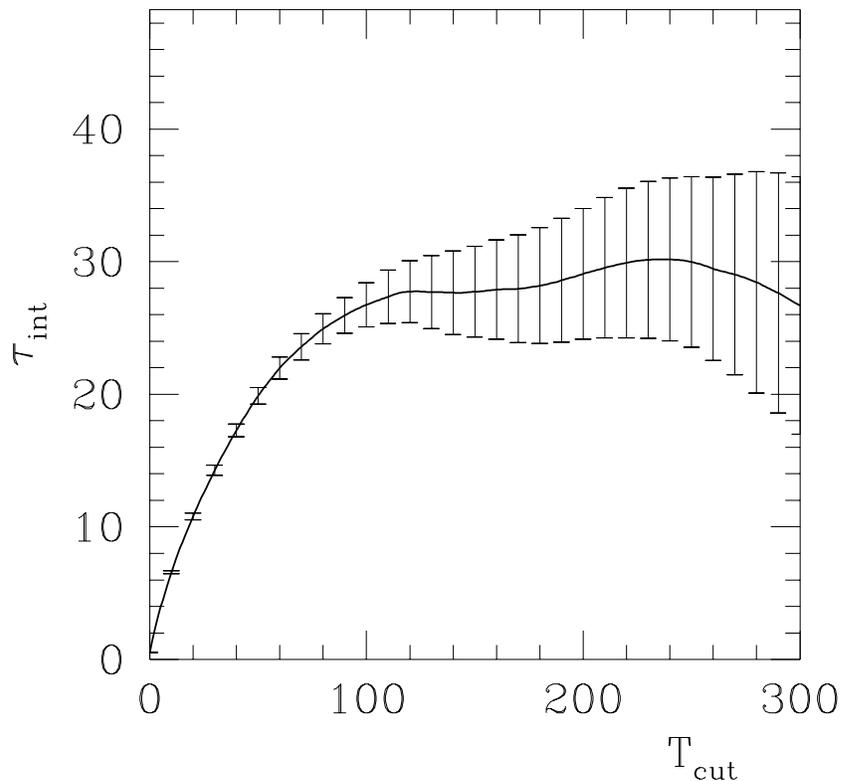

Figure 1: We plot the integrated autocorrelation function for the lowest eigenvalue $\tilde{\lambda}_0$ of $\tilde{M}^\dagger \tilde{M}$ as obtained from the Hybrid Monte Carlo algorithm on a $6^3 12$ lattice at $\beta = 2.12$, $\kappa = 0.15$.

Second, on the $12^4$ lattice, we ran the Kramers equation algorithm at two values of $\gamma$: (a) $\gamma = 0.5$ and (b) $\gamma = 2.0$. The behavior of the integrated autocorrelation function is given in Figure 2. It shows that tuning of this free parameter $\gamma$ in the Kramers equation algorithm can change the autocorrelation times by almost a factor of two. It is interesting to note that, whereas the autocorrelation times of the fermion correlators are somewhat increased, the ones for $U_p$ and $\tilde{\lambda}_0$ are decreased. We found a similar effect of the tuning of $\gamma$ in runs on $8^3 12$ lattices. Again, while increasing $\gamma$, the autocorrelation times for the fermionic correlators increased whereas the ones for $U_p$ and $\tilde{\lambda}_0$ dropped.

Table 4: Results for the autocorrelation times. The entry in the last column $\bar{\tau}_{int}(U_p)N_{CG}$ gives the product of the plaquette integrated autocorrelation time and the average number of Conjugate Gradient iterations per trajectory. For the $12^4$ lattice we ran the Kramers equation algorithm with two values of $\gamma$: (a) $\gamma = 0.5$ and (b) $\gamma = 2.0$.

| Algorithm | $L_s^3 L_t$ | $\bar{\tau}_{int}(U_p)$ | $\bar{\tau}_{int}(C_\pi)$ | $\tau_{exp}(U_p)$ | $\tau_{exp}(C_\pi)$ | $\bar{\tau}_{int}(U_p)N_{CG}$ |
|---|---|---|---|---|---|---|
| Kramers | $6^3 12$ | 54(10) | 60(20) | 60(7) | 63(9) | $10.5(1.9)\cdot 10^3$ |
| HMC | $6^3 12$ | 17(2) | 28(8) | 22(3) | 23(3) | $12.6(1.6)\cdot 10^3$ |
| Kramers (a) | $12^4$ | 64(24) | 32(16) | 68(20) | 32(5) | $14.1(5.3)\cdot 10^3$ |
| Kramers (b) | $12^4$ | 30(9) | 28(12) | 44(17) | 48(17) | $6.6(2.2)\cdot 10^3$ |
| HMC | $12^4$ | 6(1) | 6(2) | 6(1) | 5(1) | $7.0(1.1)\cdot 10^3$ |
| Kramers | $16^4$ | 70(30) | 60(20) | 88(14) | 60(14) | $13.2(5.7)\cdot 10^3$ |

# 6 Reversibility

The detailed balance proof of the HMC algorithm requires *exact* reversibility [1] of the discretized equations of motion. Of course, on a computer, the algorithm runs with only a finite precision and one may wonder, whether accumulations of rounding errors can lead to violations of the exact reversibility. This is particularly true for todays simulations where $O(100)$ molecular dynamics steps are used to obtain one trajectory [7]. One may investigate this question by taking an initial configuration $C_{ini}$, integrating it along a trajectory and then integrating it back to reach the end configuration $C_{end}$. Indeed, measuring several observables on $C_{ini}$ and $C_{end}$ discrepancies were encountered [8].

We decided to study this problem in numerical simulations using the HMC algorithm. The nonlinear nature of Hamilton's equations of motion that are used in the HMC algorithm suggests that the dynamics is chaotic. To test this proposal, let us define a quantity, $\|dU\|$, to measure the difference between two gauge field configurations

$$\|dU\|^2 = \frac{1}{4\Omega} \sum_{x,\mu,a} (U_\mu^a(x) - V_\mu^a(x))^2 \ . \tag{23}$$

Here $U_\mu^a(x), V_\mu^a(x)$ are two SU(2) gauge link variables with lattice point index $x$, direction $\mu$ and group index $a$.

To show that the suggested chaotic behavior is really a property of Hamilton's equations of motion, we proceeded in the following way. Given a gauge field configuration obtained in the course of some run, we added a small noise $\delta U_\mu(x)$ to the gauge field variable $U_\mu(x)$, such that $V_\mu(x) = U_\mu(x) + \delta U_\mu(x)$. Then, we took both configurations and iterated them according to the leapfrog integration scheme used in the HMC algorithm. We measured $\|dU\|$ after some number of steps $N_{md}$ in the leapfrog integration. If the system is chaotic, we expect that asymptotically ($\epsilon N_{md} \gg 1$)

$$\|dU\| = A e^{\nu \epsilon N_{md}}. \tag{24}$$

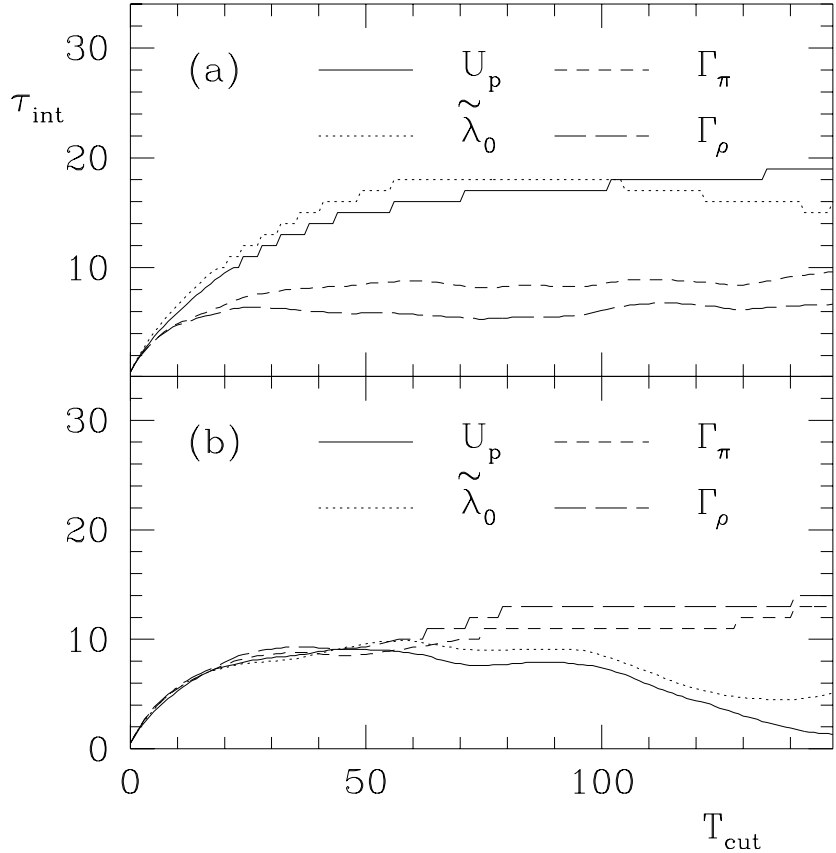

Figure 2: Integrated autocorrelation function on the $12^4$ lattice for two values of $\gamma$, (a) $\gamma = 0.5$, (b) $\gamma = 2.0$. The integrated autocorrelation function is plotted in units of measurements, which we performed every fourth trajectory for this run.

In eq. (24) $\nu$ is the –to be determined– Liapunov exponent, characterizing a chaotic system and $\epsilon$ is the step size used in the program.

In Fig. 3a, we show $lg(\|dU\|)$ as a function of $\epsilon N_{md}$, where $lg$ stands for the 10-based logarithm. Clearly, an asymptotic linear behavior is seen, giving a Liapunov exponent of $\nu = 0.78$. The step size $\epsilon$ was chosen to be $\epsilon = 0.01$ and the lattice is $\Omega = 4^4$. The data are obtained by averaging over 10 independent gauge field configurations. Errors are smaller than the symbol size.

Obviously, the noise that has been added is faking some rounding errors that occur in the actual simulations. *If* such rounding errors appear, we predict then that the errors blow up

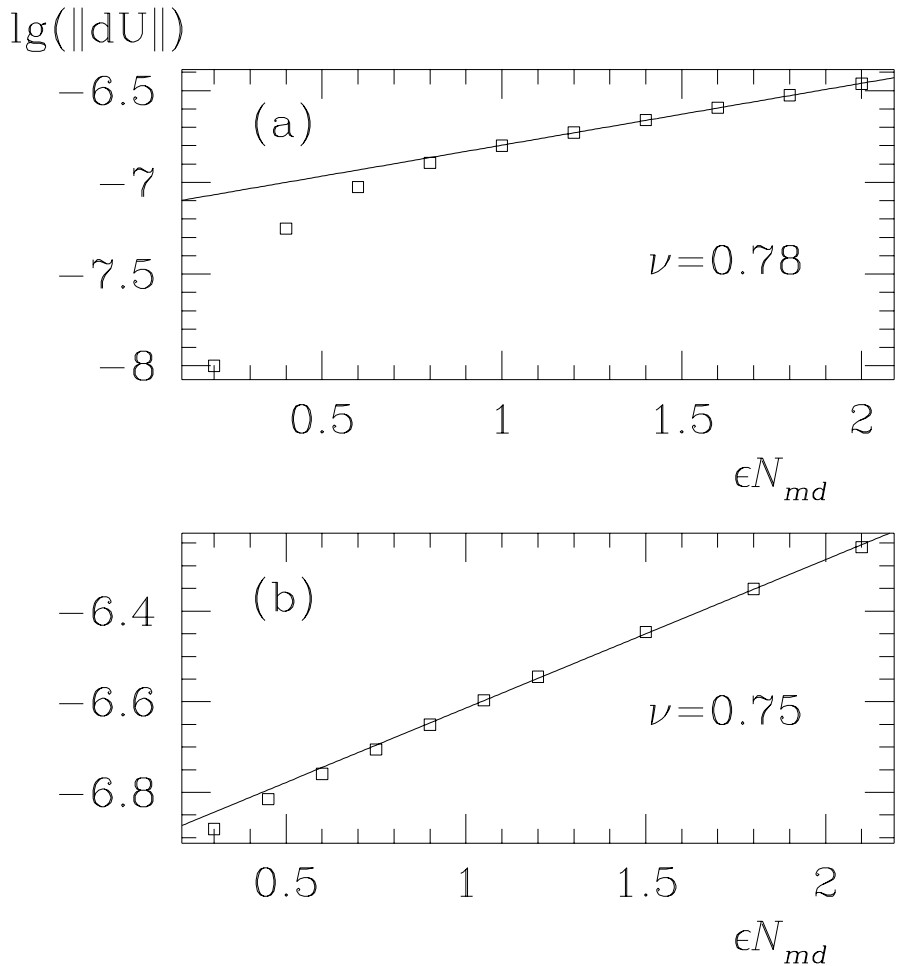

Figure 3: We plot $lg(\|dU\|)$ as defined in eq. (23). In (a) $\|dU\|$ is obtained by adding a small noise $\delta U$ to the initial configuration and then iterate the leapfrog integration. In (b) $\|dU\|$ is obtained by reversing the trajectory.

exponentially according to eq. (24) with the Liapunov exponent as found above.

To see whether this effect really happens, we performed a Monte Carlo run with our dynamical Wilson fermion program as described above on $4^4$ lattices. Every third trajectory, we reversed the time and integrated back, measuring $\|dU\|$ using the initial configuration before starting the leapfrog integration and the final configuration at the end of the reversed trajectory. By fixing the step size $\epsilon = 0.03$, we obtained $\|dU\|$ as a function of the trajectory length $\epsilon N_{md}$. Indeed, we find in Fig. 3b again a linear behavior of $lg(\|dU\|)$ with a Liapunov exponent that is compatible with the previous one.

These figures were obtained from runs on workstations with 32 bit arithmetic. We repeated the runs with 64 bit arithmetic. We found the same asymtotic exponential behavior with the only difference that the amplitude $A$ in eq. (24) gets smaller and the asymptotic behavior sets in much later. We also measured the Liapunov exponent on the APE machine which has 32 bit arithmetic but where we used double precision or Kahan summations [20, 21] to avoid rounding errors in scalar products. Again, we find a Liapunov exponent around $\nu = 0.75$. Notice that, for smaller trajectory lengths, the value of $\|dU\|$ is less than what the asymptotic formula (24) predicts. For example, for a $4^4$ lattice, when comparing the value of $\|dU\|$ at trajectory length one (typical for HMC algorithm) and at trajectory length 0.1 (typical for the Kramers equation algorithm), we found a factor of 2.5 difference. This factor will increase further for larger lattices. From this point of view, the Kramers equation algorithm appears to be a safer algorithm than the HMC algorithm.

To see whether the Liapunov exponent is really representing the chaotic behavior in the continuum time, we tried to extract it for different values of the integration step size $\epsilon$. We find that $\nu$ is independent of the step size, provided the step size is small enough. Therefore, we expect the Liapunov exponent to represent the property of the continuum time integration.

We have also investigated the dependence of the Liapunov exponent $\nu$ on bare parameters. On the $4^4$ lattice, we have performed the study of the quantity $\|dU\|$ at various $\kappa$ values for $\beta = 1.75$. We found that the Liapunov exponent $\nu$ has very weak dependence on $\kappa$. Runs on a larger $6^3 12$ lattice revealed again a value of $\nu \sim 0.75$, strengthening our conclusion that it is representing the continuum time chaotic behavior of the classical equations of motion.

# 7 Conclusion

In this paper, we compared the performance of the HMC and the Kramers equation algorithms in simulations of QCD with two flavors of dynamical Wilson fermions and gauge group SU(2). Adopting the same improvements for both algorithms, preconditioning and the Sexton-Weingarten [9] leapfrog integration scheme, we found that the Kramers equation algorithm performs as well as the HMC algorithm, as can be read off from table 4. Although the integrated autocorrelation time is larger for the Kramers equation algorithm (see table 4), the average number of Conjugate Gradient iterations per trajectory is much less, resulting in a comparable performance for the Kramers equation algorithm. Therefore, we think that the Kramers equation algorithm should be used more often in QCD simulations so as to obtain more experience with it. In particular, the tuning of the parameter $\gamma$ in the Kramers equation algorithm seems to be important to obtain optimal performance for larger lattices. As the description of the Kramers equation algorithm in section 3 shows, it is easy to change an existing HMC code to a Kramers equation one.

We demonstrated that Hamilton's equations of motion used in the HMC algorithm, due to their nonlinear nature, behave like those of a chaotic dynamical system. Therefore, *if* rounding errors occur during a simulation, they accumulate and blow up exponentially $\propto \exp\{\nu \epsilon N_{md}\}$ with $\epsilon N_{md}$ the trajectory length and $\nu$ the Liapunov exponent which we determined to be $\nu \sim 0.75$. Therefore, the HMC algorithm may lack the reversibility property. Since in the Kramers equation algorithm, a trajectory consists of only one step, this effect is substantially reduced.

## Acknowledgements

We thank M. Lüscher for essential comments and proposals during the course of this work. We are indebted to R. Edwards and A. Kennedy for a very helpful correspondence and providing nice color plots demonstrating the lack of reversibility in the HMC algorithm. All numerical simulation results have been obtained on the Alenia Quadrics (APE) computers at DESY.

# A  Appendix: Stability of the Kramers equation algorithm

For any algorithm it is sufficient to show that the transition probability P satisfies the *irreducibility* and the *stationarity* condition in order to guarantee the convergence to the equilibrium distribution [13]. Due to the refreshment of the momenta the algorithm is ergodic as is the HMC algorithm. Here we show that also the stability condition is fulfilled for the Kramers equation algorithm. In [6] it was shown that the Kramers equation algorithm even fulfills detailed balance and we therefore give the stability proof only as an additional part to make the paper self-contained.

The stability criterion reads

$$\int dz \, \pi(z) P(z \to y) = \pi(y) \,, \tag{25}$$

where we denote with $\pi(z)$ the equilibrium distribution, $z = (q,p)$ is a state in the algorithm in the field and momenta variables and $P(z \to y)$ is the transition probability, which in case of the Kramers algorithm reads

$$P(q_1, p_1 \to q_2, p_2) \tag{26}$$
$$= \int dp \, \delta(q_2 - I_Q(q_1, p_1)) \delta(p - I_P(q_1, p_1))$$
$$\min\{1, e^{H(q_1,p_1) - H(q_2,p)}\} \frac{e^{-\frac{(p_2 - xp)^2}{2(1-x^2)}}}{\sqrt{2\pi(1-x^2)}}$$
$$+ \int dp \, \delta(q_2 - q_1) \delta(p + p_1)$$
$$\left(1 - \min\{1, e^{H(q_1,p_1) - H(I_Q(q_1,p_1), I_P(q_1,p_1))}\}\right) \frac{e^{-\frac{(p_2 - xp)^2}{2(1-x^2)}}}{\sqrt{2\pi(1-x^2)}} \,. \tag{27}$$

In (26) $0 < x < 1$ is arbitrary. $I_{Q(P)}$ correspond to the leapfrog integrators as described in the text. The only important property of them is that they are linear operators with exact time-reversal symmetry. With the transition probability as given in (26), it is easy to verify the stability criterion:

$$\int dq_1 dp_1 dp \, e^{-H(q_1,p_1)} P(q_1, p_1 \to q_2, p_2) \tag{28}$$
$$= \int dq_1 dp_1 dp \, \delta(q_1 - I_Q(q_2, -p)) \delta(p_1 - I_P(q_2, -p))$$
$$\min\{e^{-H(q_1,p_1)}, e^{-H(q_2,p)}\} \frac{e^{-\frac{(p_2 - xp)^2}{2(1-x^2)}}}{\sqrt{2\pi(1-x^2)}}$$
$$+ \int dq_1 dp_1 dp \, \delta(q_2 - q_1) \delta(p + p_1)$$

$$\left(e^{-H(q_1,p_1)} - \min\{e^{-H(q_1,p_1)}, e^{-H(I_Q(q_1,p_1),I_P(q_1,p_1))}\}\right) \frac{e^{-\frac{(p_2-xp)^2}{2(1-x^2)}}}{\sqrt{2\pi(1-x^2)}}$$

$$= \int dp \min\{e^{-H(I_Q(q_2,-p),I_P(q_2,-p))}, e^{-H(q_2,p)}\} \frac{e^{-\frac{(p_2-xp)^2}{2(1-x^2)}}}{\sqrt{2\pi(1-x^2)}}$$

$$+ \int dp e^{-H(q_2,-p)} \frac{e^{-\frac{(p_2-xp)^2}{2(1-x^2)}}}{\sqrt{2\pi(1-x^2)}}$$

$$- \int dp \min\{e^{-H(q_2,-p)}, e^{-H(I_Q(q_2,-p),I_P(q_2,-p))}\} \frac{e^{-\frac{(p_2-xp)^2}{2(1-x^2)}}}{\sqrt{2\pi(1-x^2)}}, \tag{29}$$

where in the first step we have used the fact that the integrators are reversible under a sign change of the momenta. Since the terms containing the minimum conditions cancel, and the Hamiltonian is quadratic in the momenta, we arrive at

$$\int dp e^{-H(q_2,p)} \frac{e^{-\frac{(p_2-xp)^2}{2(1-x^2)}}}{\sqrt{2\pi(1-x^2)}} = e^{-H(q_2,p_2)}. \tag{30}$$

Hence the stability condition is shown.